\documentstyle[12pt,epsf]{article}

\begin{document}
\def\comment#1{\marginpar{{\scriptsize #1}}}
\def\framew#1{\fbox{#1}}                        
\def\framep#1{\noindent \fbox{\vbox{#1}}}       
\def\framef#1{\fbox{\vbox{ #1 }}}               
\def\goto{$\ra $ }

\def\be{\begin{equation}}
\def\ee{\end{equation}}
\def\bea{\begin{eqnarray}}
\def\eea{\end{eqnarray}}

\newtheorem{proposition}{Proposition}[section]
\def\bprop{\bigskip\begin{proposition}~~~\\ \rm}
\def\eprop{\end{proposition}\bigskip}
\def\proof{\bigskip \noindent {\it Proof.} \ }
\newtheorem{naming}{Definition}[section]   
\def\bnam{\bigskip\begin{naming}~~~\\ \rm}
\def\enam{end{naming}\bigskip}

\bibliographystyle{unsrt}
\def\br{\begin{thebibliography}{abc}}
\def\er{\end{thebibliography}}
\def\rf{\bibitem}
\def\cstars{$C^*$-algebras }
\def\cstar{$C^*$-algebra }
\def\unit{I\!\!I}
\def\norm#1{\parallel #1 \parallel}
\def\abs#1{\left| #1\right|}
\def\ha{\widehat{\cal A}}
\def\hc{\widehat{\cal C}}
\def\prim{$Prim \ca~$}

\def\bar#1{\overline{#1}}
\def\what{\widehat}
\def\wtilde{\wtilde}
\def\sp{~~~~~}
\def\bra#1{\left\langle #1\right|}
\def\ket#1{\left| #1\right\rangle}
\def\EV#1#2{\left\langle #1\vert #2\right\rangle}
\def\VEV#1{\left\langle #1\right\rangle}
\def\pa{\partial}
\def\del{\nabla}

\def\a{\alpha}
\def\b{\beta}
\def\c{\raisebox{.4ex}{$\chi$}}
\def\d{\delta}
\def\e{\epsilon}
\def\f{\phi}
\def\g{\gamma}
\def\h{\eta}
\def\i{\iota}
\def\j{\psi}
\def\k{\kappa}
\def\l{\lambda}
\def\m{\mu}
\def\n{\nu}
\def\o{\omega}
\def\p{\pi}
\def\q{\theta}
\def\r{\rho}
\def\s{\sigma}
\def\t{\tau}
\def\u{\upsilon}
\def\x{\xi}
\def\z{\zeta}

\def\D{\Delta}
\def\F{\Phi}
\def\G{\Gamma}
\def\J{\Psi}
\def\L{\Lambda}
\def\O{\Omega}
\def\P{\Pi}
\def\Q{\Theta}
\def\S{\Sigma}
\def\U{\Upsilon}
\def\X{\Xi}
\def\Z{\Zeta}

\def\ca{{\cal A}}
\def\cb{{\cal B}}
\def\cc{{\cal C}}
\def\cd{{\cal D}}
\def\ce{{\cal E}}
\def\cf{{\cal F}}
\def\cg{{\cal G}}
\def\ch{{\cal H}}
\def\ci{{\cal I}}
\def\cj{{\cal J}}
\def\ck{{\cal K}}
\def\cl{{\cal L}}
\def\cm{{\cal M}}
\def\cn{{\cal N}}
\def\co{{\cal O}}
\def\cp{{\cal P}}
\def\cq{{\cal Q}}
\def\czr{{\cal R}}
\def\cs{{\cal S}}
\def\ct{{\cal T}}
\def\cu{{\cal U}}
\def\cv{{\cal V}}
\def\cw{{\cal W}}
\def\cx{{\cal X}}
\def\cy{{\cal }}
\def\cz{{\cal Z}}
%
%
\def\inbar{\,\vrule height1.5ex width.4pt depth0pt}
\def\IG{\relax\,\hbox{$\inbar\kern-.3em{\rm G}$}}
\def\IU{\relax\,\hbox{$\inbar\kern-.3em{\rm U}$}}
\def\ID{\relax{\rm I\kern-.18em D}}
\def\IF{\relax{\rm I\kern-.18em F}}
\def\IH{\relax{\rm I\kern-.18em H}}
\def\II{\relax{\rm I\kern-.17em I}}
\def\I1{\relax{\rm 1\kern-.28em l}}
\def\IM{\relax{\rm I\kern-.18em M}}
\def\IN{\relax{\rm I\kern-.18em N}}
\def\IP{\relax{\rm I\kern-.18em P}}
\def\IQ{\relax\,\hbox{$\inbar\kern-.3em{\rm Q}$}}

\def\IC{\hbox{{\bf \inbar}\hskip-4.0pt C}}

\def\IR{\hbox{I\hskip-1.7pt R}}
\font\cmss=cmss10 \font\cmsss=cmss10 at 7pt
\def\IZ{\relax\ifmmode\mathchoice
{\hbox{\cmss Z\kern-.4em Z}}{\hbox{\cmss Z\kern-.4em Z}}
{\lower.9pt\hbox{\cmsss Z\kern-.4em Z}}
{\lower1.2pt\hbox{\cmsss Z\kern-.4em Z}}\else{\cmss Z\kern-.4emZ}\fi}
%
\def\Up{\Uparrow}
\def\up{\uparrow}
\def\Dn{\Downarrow}
\def\dn{\downarrow}
\def\Rt{\Rightarrow}
\def\rt{\rightarrow}
\def\Lt{\Leftarrow}
\def\lt{\leftarrow}
\def\bc{{\bf{C}}}

\newcommand{\paf}[2]{\frac{\partial#1}{\partial#2}}
\renewcommand{\thefootnote}{\fnsymbol{footnote}}
\def\fn{\footnote}
\footskip 1.0cm

\thispagestyle{empty}
\setcounter{page}{0}

\hfill \today

\vskip2cm

\centerline {\Large {\bf A NEW PHASE DIAGRAM}}
\vspace{5mm}
\centerline {\Large {\bf FOR THE t-J MODEL}}
\vspace{5mm}
\centerline{\Large  {\bf{}}}
\vspace{0.75cm}
\centerline {Elisa Ercolessi$^{a)}$,
             Pierbiagio Pieri$^{a,b)}$, 
             Marco Roncaglia$^{a,b)}$ }
\vspace{1cm}
\centerline {\it $^{a)}$Dipartimento di Fisica, Universit\`a di
Bologna and INFM }
\centerline {\it Via Irnerio 46, I-40126, Bologna, Italy.}
\centerline {\it $^{b)}$INFN, Sezione di Bologna, Italy.}

\vspace{.5cm}

\begin{abstract}
We study the so-called nonmagnetic phases (dimer and flux states) in
the t-J model below half filling. We present a new phase
diagram, at zero and finite temperature, that includes broad
areas of phase coexistence (dimer-flux or flux-uniform), in
accordance with experimental and numerical data on the possibility of
separation into hole-rich and hole-poor regions. We also briefly
comment on some techniques used in the literature to discuss phase
separation in the t-J model.

\end{abstract}
\newpage

\setcounter{footnote}{0}

Since Anderson's suggestion \cite{A2} that strongly correlated electron
models might be relevant to describe the physics of high-T$_c$
superconductors, much work, both numerical and analytical, has been
done in order to understand the properties of the Hubbard model in
2-dimensions, in particular by examining simplified models
that can be obtained from it in the strong coupling limit, such as
the Heisenberg model or the t-J model \cite{GJR} for the one-band
version  or the spin-fermion model \cite{MZ} for the three-band case.

Still, the question of the ground state of these models has not been
settled, at least away from half-filling. As candidate for the ground
state, a variety of phases has been proposed, which include the
long-range antiferromagnetically ordered (AF) state \cite{Hi}, the
short-range antiferromagnetically ordered (RVB) state \cite{A1}, as
well as the so called non-magnetic states, such as the dimer and the
flux phase \cite{AM}.

In this paper, we will concentrate on the t-J model, which has been
extensively studied by means of different techniques and within
different approaches, that include standard mean field theory
[6-8], slave-boson techniques [9-12] and 
numerical simulations \cite{NUM,DR}. 

At half-filling, i.e. at zero doping ($\d=0$), when the model reduces to
the well known spin 1/2 Heisenberg Hamiltonian, the instability of the
other phases against the AF phase has been established beyond any
reasonable doubt by analytical and numerical methods \cite{HG}.
Within mean field theory, the lowest energy phase is not the
AF but the dimer phase, either the columnar or the staggered one 
\cite{DEMT,RS}
which are degenerate at this level of approximation. Even the inclusion
of both classical and quantum quadratic fluctuations does not increase
the energy of the dimer phase above the energy of the AF state
\cite{DEMTV}. It is known that this apparent contradiction is strictly
related to the difficulty of taking into account the constraint of one
electron per site in this approach. Indeed, in \cite{DEMTV} it has been
shown that at the mean field level, eventually corrected by quadratic
fluctuations,  the constraint is implemented exactly for the AF phase
but not for the dimer phase. Thus, in the latter case we are
effectively dealing with a larger space of states and hence it comes of
no surprise the fact the energy gets substantially lowered.

For finite doping fraction ($\d> 0$), the major difficulty consists of
developing good approximating techniques while dealing with the
Gutzwiller (or below-half-filling) constraint. The experimental
evidence and all the above mentioned theoretical studies 
confirm the intuitive idea that the presence of holes tends to
disfavour a long-range antiferromagnetically ordered phase. However,
there is still no complete agreement on the possible ground state at
$\d \neq 0$. 
Among the candidates for the ground state of the t-J model, 
the RVB and the flux states have been mainly considered.  
The former, first proposed by Anderson and coworkers \cite{BZA}, is 
characterized by a superconducting ordered parameter of the kind 
$\D_{ij} = \langle c_{i\uparrow} c_{j\downarrow} - c_{i\downarrow} 
c_{j\uparrow} \rangle $,
where $i,j$ are nearest neighborhood sites, which gives short-range
antiferromagnetic correlations. The latter \cite{AM} corresponds to a
nonmagnetic complex order parameter $U_{ij} = \langle  \sum_{\a = \uparrow,
\downarrow} c_{i\a}^\dagger c_{j\a} \rangle $, whose phase originates a
nonzero magnetic flux threading the elementary (square) plaquettes of
the lattice in a staggered way. Whenever the flux is different from
$0$ or $\p$ (mod $2\p$), this phase breaks parity and time-reversal
symmetries. This has been the starting point of a large number of
works, in which it has been argued that the continuum limit
effective action describing the low-energy excitations
around the flux phase could contain a Chern-Simons term \cite{CS}, which
would impart fractional statistics to the quasiparticles, leading to
the possibility of anyonic superconductivity \cite{ANY}. An analysis of
this claim requires more sophisticated techniques than the
one used in this paper and goes beyond our present scope. We remark
here only that, while it exists for the AF phase \cite{CON,MZ}, a
detailed study of the continuum limit around the flux phase,
able to prove or disprove the presence of a topological term, is still
missing in the literature. We plan to further discuss this
point in some future work.

At half filling, the large-U limit of the Hubbard model admits an
SU(2) gauge symmetry \cite{AZHA}, which shows the equivalence between
the apparently different RVB and flux states. This symmetry is broken
at $\d \neq 0$, so that these two states evolve into different phases,
which nevertheless stay very close in energies. In this paper we will
not address the problem of the competition between the RVB and
the flux phase \cite{SSY,COMP}. We will concentrate our analysis on the
nonmagnetic phases only. This is because we seem to find a phase
diagram in the temperature-doping (T-$\d$) plane richer than the one
considered in previous works \cite{AM} within the same kind of mean
field approximation. If, on one side, our pure mean field data confirm 
the hypothesis
that the flux phase gives the ground state in the range of the doping
fraction $\d$ which is relevant for superconductivity, on the other
side, we discover that, especially at low temperatures, there are wide
regions of phase coexistence, corresponding to either dimer-flux or
flux-uniform phase separation, depending on the value of
$\d$. This seems to go in the same direction of some 
experimental data on high temperature superconducting
materials \cite{HUN,TRA} and as weel as of some perturbative and
numerical studies of the t-J model [26-30], which find a tendency 
towards a separation of regions with different hole concentrations. 
We remark here that the presence of phase coexistence
regions might also affect a stability analysis of the RVB vs. the flux
phase. 

Our starting point is the t-J Hamiltonian, which can be obtained
as the strong coupling limit, $|t|\ll U$, of the one-band version of the
Hubbard model for small values of the doping $\d$ \cite{GJR}:
\bea 
\ch_{eff} &=& \sum_{<ij>,\a} t_{ij} (1-n_{i\bar{\a}})\,
c^\dagger_{i\a} c_{j\a} \,(1-n_{j\bar{\a}}) \label{prohub} \\
&+& \sum_{<ij>} J_{ij} \left\{ \vec{S}_i \cdot \vec{S}_j - \frac{1}{4} 
n_{i} n_{j} \right\} \nonumber 
\eea
where $c^\dagger_{i\a}$ is the creation operator of one electron with
spin $\a=\uparrow, \downarrow$ in the site $i$ and 
$\vec{S}_i= \frac{1}{2} \sum_{\a\b} c^\dagger_{i\a} \vec{\s}_{\a\b}
c_{i\b}$ ($\vec{\s}=(\s_1,\s_2,\s_3)$ being Pauli matrices) are the
spin operators on the site $i$. In addition, $J_{ij} = \frac{4
|t_{ij}|^2}{U}$, $\bar{\a} = -\a$ while the symbol $\langle ij\rangle $ 
denotes a sum
over nearest-neighbor (n.n.) lattice sites only.  The operator
expressions containing $n_{i\a}\equiv c^\dagger_{i\a} c_{i\a}$  have
the effect of enforcing the Gutzwiller projector.

At exact half-filling, when $n_i \equiv n_{i\uparrow} +
n_{i\downarrow} = 1$, the first  term of
(\ref{prohub}) vanishes identically, while the second line reduces to
the  spin $\frac{1}{2}$ antiferromagnetic Heisenberg
Hamiltonian. If we introduce holes in the system, so that the doping
fraction is $\d \neq 0$, we can no longer neglect the first term of
(\ref{prohub}) which describes a direct hopping of one electron from
the site $j$ to the (empty) n.n. site $i$.
Because of the presence of the number operators $n_{i\a}$, the
hopping term is the sum of products of up to six
electron creation/annihilation operators and is therefore very
difficult to analyze.

Thus, following  Anderson and coworkers \cite{BZA}, we make the
assumption that the main effect of the Gutzwiller projection  is the
renormalization of the hopping amplitude from its nominal
value $t_{ij}$ to $\d t_{ij}$. We recall that such assumption has been
proven to be correct either within a slave boson approach at the mean
field level \cite{KL,GK} and within a variational approach making use
of the Gutzwiller approximation \cite{S}.
With the further assumption that
$t_{ij}=t$, up to nonrelevant constant factors, the Hamiltonian 
(\ref{prohub}) becomes:  
\be
\ch = t\d \sum_{<ij>} \sum_{\a} c^\dagger_{i\a} c_{j\a}
- \frac{J}{2} \sum_{(ij)} \sum_{\a\b} c^\dagger_{i\a} c_{j\a}
c^\dagger_{j\b} c_{i\b} \; . \label{tj}
\ee
We will also restrict our attention to 2-D square lattices.

To study the phase diagram of (\ref{tj}), we will follow the
technique described in \cite{DEMT} to rewrite the  partition function in
the grand canonical ensemble 
$\cz = Tr \left\{ e^{-\b\, (\ch - \m N)} \right\} $ as:
\bea
\cz &=& \int [\cd \j^*_{i\a} \, \cd \j_{i\a} ] \exp \left\{ 
- \int_0^1 d\t \left[ \sum_i \sum_{\a} \j^*_{i\a} (\partial_{\t} - 
\m \b) \j_{i\a} \right. \right. \label{pfun} \\
&+& \left. \left. \b t \d \sum_{<ij>} \sum_{\a} \j^*_{i\a} \j_{j\a} - 
\frac{\b J}{2}  \sum_{<ij>} \sum_{\a\b} \j^*_{i\a} \j_{j\a} \j^*_{j\b}
\j_{i\b} \right]  \right\}  \; ,\nonumber
\eea
where we have introduced Grassmann fields $\j_{i\a}$ for the
fermionic operators $c_{i\a}$.

We can now decouple the quartic term in the exponential of
(\ref{pfun}) via Hubbard-Stratonovich auxiliary fields
$\cu_{ij}$ as follows: 
\bea
\cz &=& \int [\cd \j^*_{i\a}\, \cd \j_{i\a} ] \int [\cd \cu^*_{ij}
\cd \cu_{ij} ] \exp\left\{ - \frac{2 \b}{J} \int_0^1 d\t  \sum_{<ij>}
\cu^*_{ij}  \cu_{ij} \right\} \label{hs} \\
&\times& \exp \left\{ - \int_0^1 d\t \left[ \sum_i \sum_{\a}\j^*_{i\a}
 (\partial_{\t} - \m \b) \j_{i\a}   \right. \right.\nonumber \\
&+&\left. \left. \b t \d \sum_{<ij>} \sum_{\a} \j^*_{i\a}
\j_{j\a} + \b \sum_{<ij>} \sum_{\b} \cu^*_{ij}
\j^*_{i\b} \j_{j\b} \right] \right\} \; . \nonumber
\eea

We will work in the static approximation: $\cu_{ij}(\t) = \cu_{ij}$
(constant in the imaginary time $\t$), so that, going to
Fourier transform with respect to $\t$, (\ref{hs}) can be rewritten as:
\bea
\cz &=& \int [\cd \cu^*_{ij} \cd \cu_{ij}] \exp\left\{ - \frac{2 \b}{J}
\sum_{<ij>} \cu^*_{ij} \cu_{ij} \right\} \int [\cd \j^*_{i\a}\, \cd
\j_{i\a}]\label{sta} \\ 
&\times&  \exp \left\{ \sum_{nn'}\sum_{ij} \sum_{\a}
\j^*_{i\a}(\o_n) \, [\IG^{-1} - \b \IU]_{nn'}^{ij} 
\j_{j\a}(\o_{n'})  \right\} \; , \nonumber
\eea
where $\o_n = (2n+1) \p$ ($n \in \IZ$) are Matsubara frequencies and
\bea
[\IG^{-1}]_{nn'}^{ij} &= &( i\o_n + \m \b )\, \d_{ij} \d_{nn'} \\
\left[ \IU\right]_{nn'}^{ij} &=& \left\{  
\begin{array}{cl} 
[t\d + \cu_{ij}] \d_{nn'} & \mbox{if $i$ n.n. $j$} \\
0 & \mbox{otherwise} \end{array} \right. \; .
\eea

We can now perform the integral over the fermionic variables to get:
\bea
\cz &=& \int [\cd \cu^*_{ij} \cd \cu_{ij}] \exp\left\{ - \frac{2 \b}{J}
\sum_{(ij)} \cu^*_{ij} \cu_{ij} \right\} \exp\left\{ -
\cs_{eff}\right\} \\ 
\cs_{eff}&\equiv& - 2 \mbox{Tr} \left\{ \log \left[ -
\IG^{-1} + \b \IU \right] \right\} \; ,
\eea
where the factor 2 comes from spin summation and ``Tr" stands for a
trace over lattice sites and frequencies.

Following \cite{AM}, we consider solutions of the saddle point
equations, $\frac{\partial \cs_{eff}}{\partial \cu_{ij}} =0$, that
admit a symmetry for translations along the diagonal of elementary
plaquette of the square lattice. In this case, the
matrix $\IU$ depends  only on  four independent link variables $\cu_j$
as shown in figure \ref{fi:lattice}(a). Under this assumption, the
corresponding Brillouin zone gets  halved. The Reduced Brillouin Zone
(RBZ) is given by the shaded area of figure \ref{fi:lattice}(b).

\begin{figure}[htb]
\begin{center}
\begin{picture}(400,200)(70,100)

\put(150,250){\circle*{5}}
\put(200,250){\circle{5}}
\put(250,250){\circle*{5}}
\put(150,200){\circle{5}}
\put(200,200){\circle*{5}}
\put(250,200){\circle{5}}
\put(150,150){\circle*{5}}
\put(200,150){\circle{5}}
\put(250,150){\circle*{5}}

\put(150,250){\line(1,0){50}}
\put(195,250){\line(-1,1){10}}
\put(195,250){\line(-1,-1){10}}

\put(200,250){\line(0,-1){50}}
\put(200,205){\line(1,1){10}}
\put(200,205){\line(-1,1){10}}

\put(250,250){\line(-1,0){50}}
\put(205,250){\line(1,1){10}}
\put(205,250){\line(1,-1){10}}

\put(150,200){\line(0,-1){50}}
\put(150,155){\line(1,1){10}}
\put(150,155){\line(-1,1){10}}

\put(150,200){\line(0,1){50}}
\put(150,245){\line(-1,-1){10}}
\put(150,245){\line(1,-1){10}}

\put(200,200){\line(1,0){50}}
\put(245,200){\line(-1,1){10}}
\put(245,200){\line(-1,-1){10}}

\put(200,200){\line(-1,0){50}}
\put(155,200){\line(1,1){10}}
\put(155,200){\line(1,-1){10}}

\put(250,200){\line(0,-1){50}}
\put(250,155){\line(1,1){10}}
\put(250,155){\line(-1,1){10}}

\put(250,200){\line(0,1){50}}
\put(250,245){\line(1,-1){10}}
\put(250,245){\line(-1,-1){10}}

\put(150,150){\line(1,0){50}}
\put(195,150){\line(-1,1){10}}
\put(195,150){\line(-1,-1){10}}

\put(200,150){\line(0,1){50}}
\put(200,195){\line(1,-1){10}}
\put(200,195){\line(-1,-1){10}}

\put(250,150){\line(-1,0){50}}
\put(205,150){\line(1,-1){10}}
\put(205,150){\line(1,1){10}}

\put(170,253){$\cu_2$}
\put(220,253){$\cu_4$}
\put(170,203){$\cu_4$}
\put(220,203){$\cu_2$}
\put(170,153){$\cu_2$}
\put(220,153){$\cu_4$}
\put(138,220){$\cu_3$}
\put(138,170){$\cu_1$}
\put(188,220){$\cu_1$}
\put(188,170){$\cu_3$}
\put(238,220){$\cu_3$}
\put(238,170){$\cu_1$}

\put(192,120){(a)}

\put(300,200){\vector(1,0){120}}
\put(360,140){\vector(0,1){120}}
\put(425,198){$k_x$}
\put(358,265){$k_y$}
\put(320,160){\dashbox{.5}(80,80)}
\put(320,200){\line(1,-1){40}}
\put(360,240){\line(1,-1){40}}
\put(340,220){\line(1,1){20}}
\put(340,220){\line(-1,-1){20}}
\put(347,213){\line(1,1){20}}
\put(347,213){\line(-1,-1){20}}
\put(354,206){\line(1,1){20}}
\put(354,206){\line(-1,-1){20}}
\put(360,200){\line(1,1){20}}
\put(360,200){\line(-1,-1){20}}
\put(366,194){\line(1,1){20}}
\put(366,194){\line(-1,-1){20}}
\put(373,187){\line(1,1){20}}
\put(373,187){\line(-1,-1){20}}
\put(380,180){\line(1,1){20}}
\put(380,180){\line(-1,-1){20}}
\put(303,193){$-\p$}
\put(402,193){$\p$}
\put(362,242){$\p$}
\put(362,153){$-\p$}

\put(352,120){(b)}

\end{picture}
\end{center}
\caption{\label{fi:lattice}}
\centerline{{\footnotesize  (a) The periodicity of the four
independent link variables $\cu_{j}$, $j=1,\cdots,4$.}}
\centerline{{\footnotesize  (b) The hatched area shows the Reduced
Brillouin Zone.}}
\vskip1cm
\end{figure}

In momentum space, the matrix $\IU$ can be easily diagonalized. It
has eigenvalues: 
\bea E_k &=& \pm \,\, |\l_k| \label{eigen}\\ 
\l_k &\equiv& \chi_1
e^{ik_x a}+\chi_2^* e^{-ik_y a}+\chi_3 e^{-ik_x a}+ \chi_4^*  e^{ik_y a}
\nonumber 
\eea 
with $\chi_j \equiv t\d + \cu_j$ and $k \in RBZ$, so that:
\be
S_{eff} = -2 \sum_{k\in RBZ} \sum_n \left[ \log(-i \o_n -\m \b + \b
|\l_k|) + \log(-i\o_n -\m \b - \b |\l_k|)\right] \; . \label{seff}
\ee
The sum over the Matsubara frequencies can now be performed,
yielding the following expression for the partition function in the
static approximation: 
\bea 
\cz &=& \int [\cd \cu_j^* \cd \cu_j] e^{-\b  \O}
\label{stapar} \\ 
\O & = &\frac{N}{J}
\sum_{j=1}^{4} |\cu_j |^2 
-\frac{2}{\b} \sum_{k\in RBZ} \left[ \log (1+e^{\b  (\m - |\l_k|)} ) 
+ \log (1+e^{\b  (\m + |\l_k|)} )\right] \; . \nonumber
\eea

The stationary points of $\O$ are thus given by the saddle-point
equations: 
\bea
\cu_1 &=& \frac{J}{N} \sum_{k\in RBZ} e^{-ik_x a} \frac{\l_k}{|\l_k|}
\left[ \frac{1}{ 1+e^{-\b (\m +|\l_k|)} } - 
\frac{1}{ 1+e^{-\b (\m -|\l_k|)} } \right] \, , \label{spe}\\
\cu_2 &=& \frac{J}{N} \sum_{k\in RBZ} e^{-ik_y a}
\frac{\l_k^*}{|\l_k|} \left[ \frac{1}{ 1+e^{-\b (\m +|\l_k|)} } - 
\frac{1}{ 1+e^{-\b (\m -|\l_k|)} } \right] \, , \\
\cu_3 &=& \frac{J}{N} \sum_{k\in RBZ} e^{ik_x a} \frac{\l_k}{|\l_k|}
\left[ \frac{1}{ 1+e^{-\b (\m +|\l_k|)} } - 
\frac{1}{ 1+e^{-\b (\m -|\l_k|)} } \right] \, ,\\
\cu_4 &=& \frac{J}{N} \sum_{k\in RBZ} e^{ik_y a} \frac{\l_k^*}{|\l_k|}
\left[ \frac{1}{ 1+e^{-\b (\m +|\l_k|)} } - 
\frac{1}{ 1+e^{-\b (\m -|\l_k|)} } \right] \, ,
\eea
which must be supplemented by the equation determining the chemical
potential $\m$:
\be
N (1-\d) = 2 \sum_{k \in RBZ} \left[ \frac{1}{ 1+e^{-\b (\m -|\l_k|)}}
+  \frac{1}{ 1+e^{-\b (\m +|\l_k|)} } \right] \, . \label{mu}
\ee

We have already recalled that, at half filling, the Hamiltonian
(\ref{tj}) admits an $SU(2)$ gauge symmetry. As a special case,
(\ref{tj}) is invariant under a local $U(1)$ transformation mapping
the operator $c_{j\a}$ into $e^{i\theta_j} c_{j\a}$. Such transformation
changes the phases of the link variables $\cu_{jk} = |\cu_{jk}|
e^{i\theta_{jk}}$ by 
\be
\theta_{jk} \rightarrow \theta_{jk} +
\theta_k-\theta_j \; ,\label{ph}
\ee
but keeps the sum of the phases around an
elementary plaquette (of vertices $j,k,l,m$) invariant: $
\D \theta = \theta_{jk} + \theta_{kl} +\theta_{lm} +\theta_{mj} =
{\rm cost.}$ Indeed, $\D \theta$ is a gauge invariant , hence
observable, quantity and can be thought of as the flux associated to a
magnetic field threading the plaquette. In addition, equations
(\ref{spe}-\ref{mu}) with $\d=0$ are left invariant and hence the
grand canonical potential $\O$ will admit a set of degenerate minima,
parametrized by $U(1)$ and all corresponding to the same $\D \theta$.
For $\d \neq 0$, this symmetry is explicitly broken by the hopping
term and, in general, only one specific choice of the phases
$\theta_{jk}$, which are no longer pure gauge degrees of freedom, will
correspond to a minimum of $\O$.

We have chosen to study equations (\ref{spe}-\ref{mu}) numerically for
the case of $t/J =1$, a value that well approximates the experimental
values for the hopping amplitude ($t= 1eV$) and the Hubbard repulsion
($U=5eV$). We have found the following solutions yielding (local) minima
of the grand canonical potential $\O$:\\ 

\begin{enumerate}

\item \underline{Uniform phase}.\\
This is characterized by the choice:
\be 
\cu_1 = \cu_2 = \cu_3 = \cu_4 \equiv \cu \in \IR^{+} \; . \label{uni}
\ee
This solution  exists at all temperature and is the only minimum of
$\O$ at high temperature and/or high doping. 
At all value of the doping $\d$, the total flux per plaquette is $\D
\theta=0$.

\item \underline{Dimer phase}.\\
It corresponds to solutions of the form:
\be\label{di}
\cu_1 \neq \cu_2 = \cu_3 = \cu_4 \; , \; \cu_j \in \IC  
\ee
and with 
\be
|\cu_1| \gg |\cu_2| \; .
\ee
At $\d=0$, the $U(1)$-gauge symmetry mentioned above allows to choose
the $\cu_j$ real. In addition, in this case, $\cu_2 =0$. For
finite $\d$, gauge invariance is broken and
the parameters $\cu_j$ are in general  complex. This is
indeed the case for a range of temperature $0 < T < T_1 = 0.1 J $, in
which the parameters stay real only up to some value $\d_{c_1}(T)$ and
then acquire a phase increasing with $\d$. For $T\geq T_1$, on
the contrary, the parameters are always real. In addition, for $\d >
\d_{c_2}(T)$, (\ref{di}) ceases to be a solution of the saddle point
equations. This behaviour is shown in figure \ref{fi:dimer}.

\begin{figure}[htb]
\begin{center}
\leavevmode
   \epsfxsize=0.50\textwidth\epsfbox{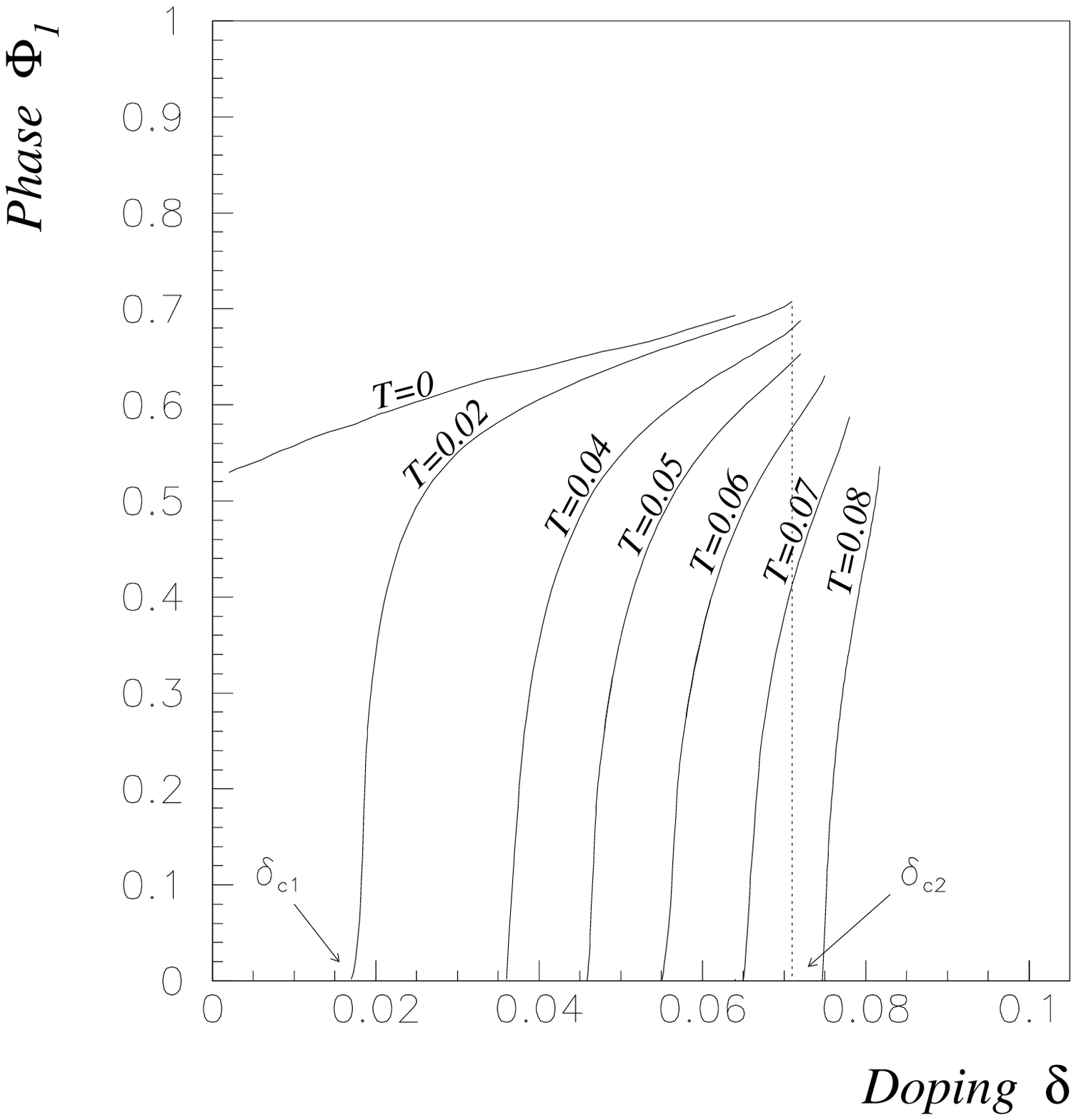}
\caption{\label{fi:dimer}}
\centerline{{\footnotesize The phase $\f_1$ of the order parameter
$U_1$ in the dimer phase}} 
\centerline{{\footnotesize as a function of the doping $\d$ at different 
temperatures.}} 
\centerline{{\footnotesize The critical dopings $\d_{c1}$ and $\d_{c1}$ 
are shown for $T=0.02$.}}
\end{center}
\end{figure}
\vskip1cm 

\item \underline{Kite phase}.\\
This solution exists only in a very small region of the parameter
space and corresponds to the choice:
\be
\cu_1 = \cu_2 \neq \cu_3 = \cu_4 \; , \; \cu_j \in \IR \; . 
\ee
The parameters $\cu_3$ and  $\cu_4$ are zero at $\d=0$ and then
increase while $\d$ increases.

\item \underline{Flux phase}.\\
This phase corresponds to a sort of complex uniform phase and is  given
by the choice:
\be
\cu_1 = \cu_2 = \cu_3 = \cu_4 \equiv \cu e^{i\f} \; , \; \cu\in \IR^+
\mbox{ and } \f \neq 0 \; .
\ee
Because of the $U(1)$ invariance, at $\d=0$, the phases of the four
parameters do not need to be the same, as long as the total magnetic
flux per plaquette is equal to $4 \f$. In figure 3, $\f$ is
shown as a function of $\d$ at different values of the temperature. For
any $T < T_c^{(0)}$, $\f$ is  $\frac{\p}{4}$ at exact half-filling
and then decreases with continuity going to zero at some
critical value of delta $\d_c(T)$. We will discuss the behaviour of the 
curve for $T=0.02$ shortly below. 
\end{enumerate}

\begin{figure}[htb]
\begin{center}
\leavevmode
   \epsfxsize=0.50\textwidth\epsfbox{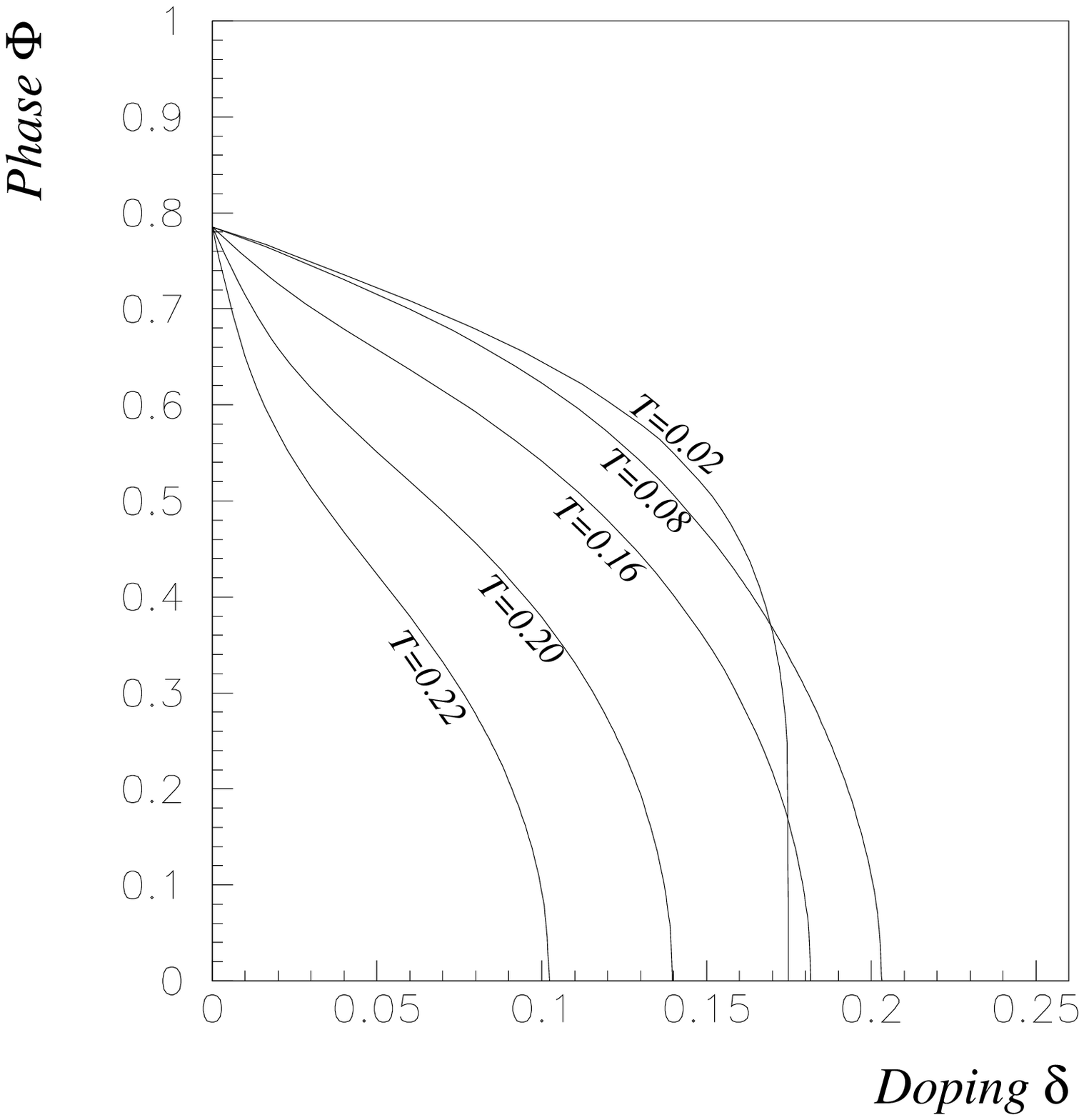}
\caption{\label{fi:flux}}
\centerline{{\footnotesize The phase $\f$ of the order parameter in
the flux phase}}
\centerline{{\footnotesize as a function of the doping $\d$ at different
temperatures.}} 
\end{center}
\end{figure}
\vskip1cm 

After having calculated and compared the free energy $F= \O
+ \m N (1-\d)$ corresponding to the different mean field solutions
described above, we have obtained the phase diagram shown in
figure 4. We remark that, contrary to what happens in
going from the dimer to the kite or the flux phase, the transition from
the flux to the uniform phase is continuous, the phases of the order
parameters $\cu_j$ going smoothly to zero as $\d$ increases.

\begin{figure}[htb]
\begin{center}
\leavevmode
   \epsfxsize=0.50\textwidth\epsfbox{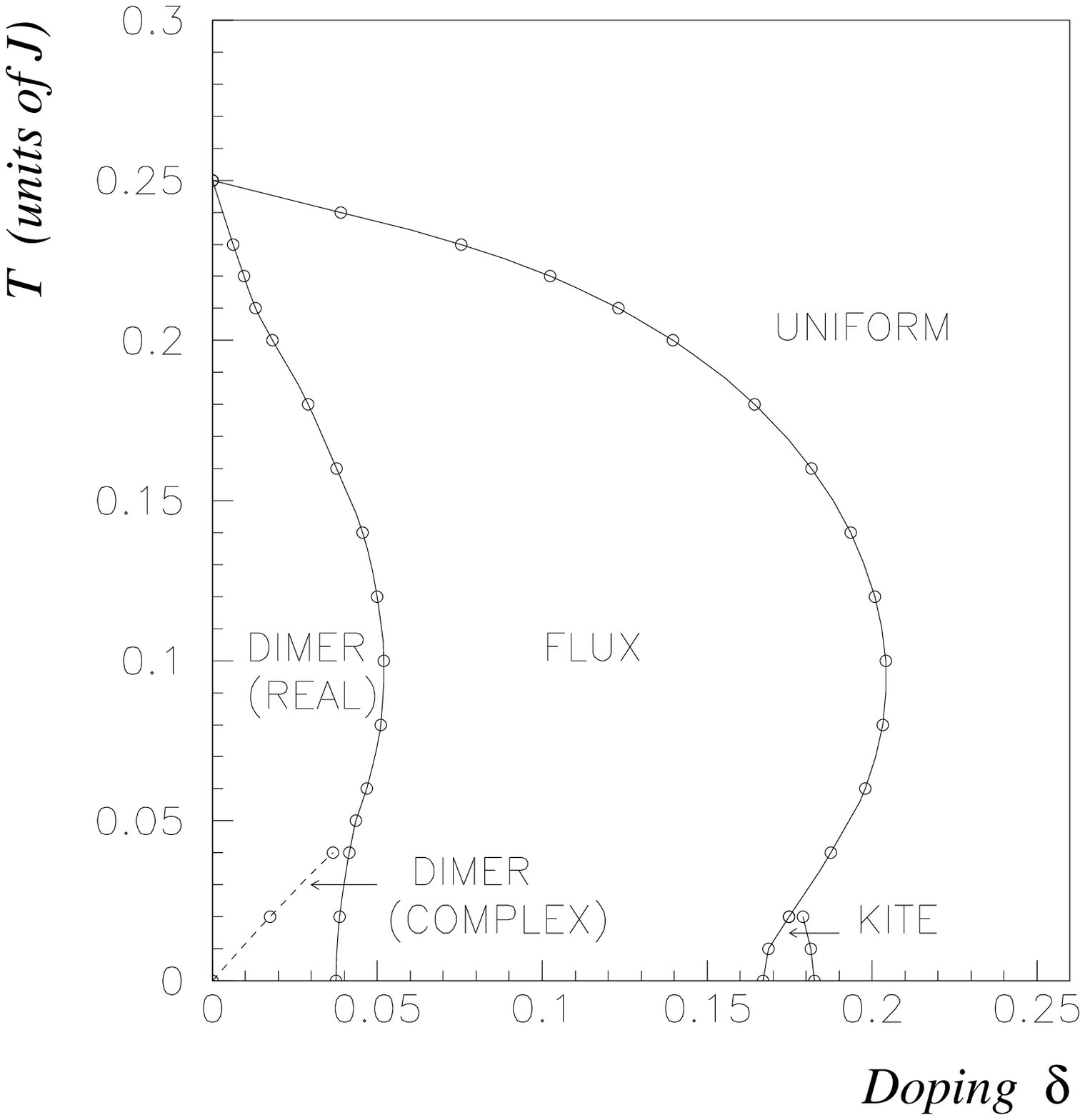}
\caption{\label{fi:meanfield}}
\centerline{{\footnotesize The phase diagram of the t-J model at
pure mean filed level.}}
\end{center}
\end{figure}
\vskip1cm 

We observe that, for $T=0$, we recover the results of \cite{AM}. 
Indeed, since we have renormalized the hopping coefficient $t$ by a 
factor of $\d$, our data have to be compared with the ones that can be
read moving  along the diagonal line 
$t/J = \d$ of the zero temperature phase diagram given in \cite{AM}. 
Also, we remark
that the flux phase has been extensively studied
\cite{SSY,UL} by analyzing the Hamiltonian (\ref{tj}) within a 
slave boson approach.  Apparently,  
some different results are obtained:
the authors of references
\cite{SSY,UL} find a flux phase corresponding to a magnetic flux per
plaquette which for small $\d$ is blocked to the value $\p$,
decreasing towards zero only after a {\it finite} value of the doping.
It is not difficult to show, however, that the Hubbard-Stratonovich
fields $\cu_{ij}'$ used in those papers differs from the $\cu_{ij}$ we
have used, the relationship being:
\bea
\cu_{ij}'& =& \cu_{ij} - \frac{t}{2J} \langle b_i^\dagger b_j\rangle  
\nonumber \\
&\simeq& \cu_{ij} - \frac{t}{2J}\; \d \label{sb} \; ,
\eea
where $b_i$ ($b_i^\dagger$) is the creation (annihilation) operator of
the slave boson and the last equality is valid in the mean field
approximation. We have checked that our numerical data for the
$\cu_{ij}$ in the flux phase, once corrected according to formula
(\ref{sb}), yield a phase for the $\cu_{ij}'$ that does indeed
reproduce the behaviour described in \cite{SSY,UL}. Thus, as it should be,
the two approaches give the same mean field results.

Our mean field calculations indicate the existence of a
rather extended region of stability for the flux phase below
half-filling, which  coincides roughly with the region relevant for
superconductivity. We notice also that, between dimer and flux as well as 
between flux and uniform phases, there is a rather strange receding of the 
phase boundary line. This is the cause of the behaviour of the $T=0.02$ 
curve in figure 3. 

The phase diagram shown
in figure 4 has been constructed without  
taking into account the possibility of phase separation. However, this
possibility cannot be overlooked. It is indeed a long-standing and 
still unresolved problem 
concerning the Hubbard model and its parent Hamiltonians, such as the t-J.   
Let us recall that experiments on different copper oxyde compounds have 
found a spatial separation between AFM hole-poor and 
superconducting hole-rich regions, happening either at a microscopic 
level \cite{HUN} or at a macroscopic level, such as in ordered striped
phases \cite{TRA}. From a theoretical point of view, much work has been
done to demonstrate that this phenomenon can occur for the Hubbard
model in the strong coupling limit. In 
particular phase separation for the t-J model has been discussed by means of 
a high-temperature expansion by Puttika et al. \cite{PU} and by Emery et al. 
\cite{EKL}
by using variational arguments and exact diagonalization numerical studies. 
Its connections with the onset of high-temperature superconductivity
has been considered in \cite{GRCDK}. Also, phase coexistence has been 
examined for other phases that might be relevant in the t-J model, such
as the spiral ones \cite{AL} or the ferromagnetic one \cite{AS,MPP},
in the latter case for its relevance to the Nagaoka problem. 
Anticipating our results, we find, for nonmagnetic phases as well, 
an instability towards phase coexistence and separation between hole-rich 
and hole-poor regions. 
Since we work in a mean field approximation, we are however not able to 
say how spatially distributed the two different regions can appear. 

We have therefore
inquired about the stability of our mean field solutions towards phase
separation by studying:\\ 
1) The fluctuation matrix for the grand canonical 
potential $\Omega$ around the saddle point solutions, since the mean
field solution is unstable whenever the fluctuation matrix has one or
more negative eigenvalues.\\
2) The behaviour of the chemical potential
$\mu$ as a function of doping $\d$, since an instability towards phase
separation is signaled by $\paf{\mu}{\d}>0$ \cite{PU}. 
In fact, if $\paf{\mu}{\d}>0$ then the isothermal compressibility 
$\kappa^{-1}=n^2\paf{\mu}{n}$ ($n=1-\d$) is negative and the system is 
clearly unstable. The stable state is given by the coexistence between 
two phases having different values of doping and the same chemical 
potential. These values of doping are given by the application of
the Maxwell construction to the diagram $\mu(\d)$ \cite{GRCDK}.\\
3) The convexity of the free energy density $f=F/V$ as a function of
$\d$, since an instability towards phase separation is indicated also
by a region of concavity of the free energy $f$. 
It is easy to see that, when working within an approximation scheme that
respects the thermodynamical equalities, the region of concavity 
of $f$ should coincide with that where $\paf{\mu}{\d}>0$, 
since $\kappa^{-1}=n^2\frac{\partial^2 f}{\partial
n^2}=n^2\paf{\mu}{n}$. In this case it is also possible to check that
the bitangent to the curve $f(\d)$ gives an interpolation between
two different phases which is equivalent to the Maxwell construction
for $\mu(\d)$. 

As for 1), we have considered only fluctuations around the mean
field solutions which preserve the symmetry under translations along the
diagonal of the square elementary plaquettes. We have found that the
fluctuation matrix develops a negative  eigenvalue in two
regions of the phase diagram: a) on the left of the boundary
separating the flux and the uniform phase, below a certain temperature
$T_c\sim 0.145 J$; b) on the left of the boundary between dimer and
flux phase, below a given temperature $T_{c1}\sim 0.085 J$. We might
therefore expect both a region of flux-uniform and a region of
dimer-flux phase coexistence.

Such scenario is fully supported by the behaviour of the chemical
potential $\mu$ as a function of $\d$. Indeed, $\paf{\mu}{\d}$ becomes
positive exactly in the regions where the fluctuation matrix develops
a negative eigenvalue. We will analyze in detail only the flux-uniform
phase coexistence, the discussion for the dimer-flux case
being completely analogous.

\begin{figure}[htb]
\begin{center}
\leavevmode
   \epsfxsize=0.50\textwidth\epsfbox{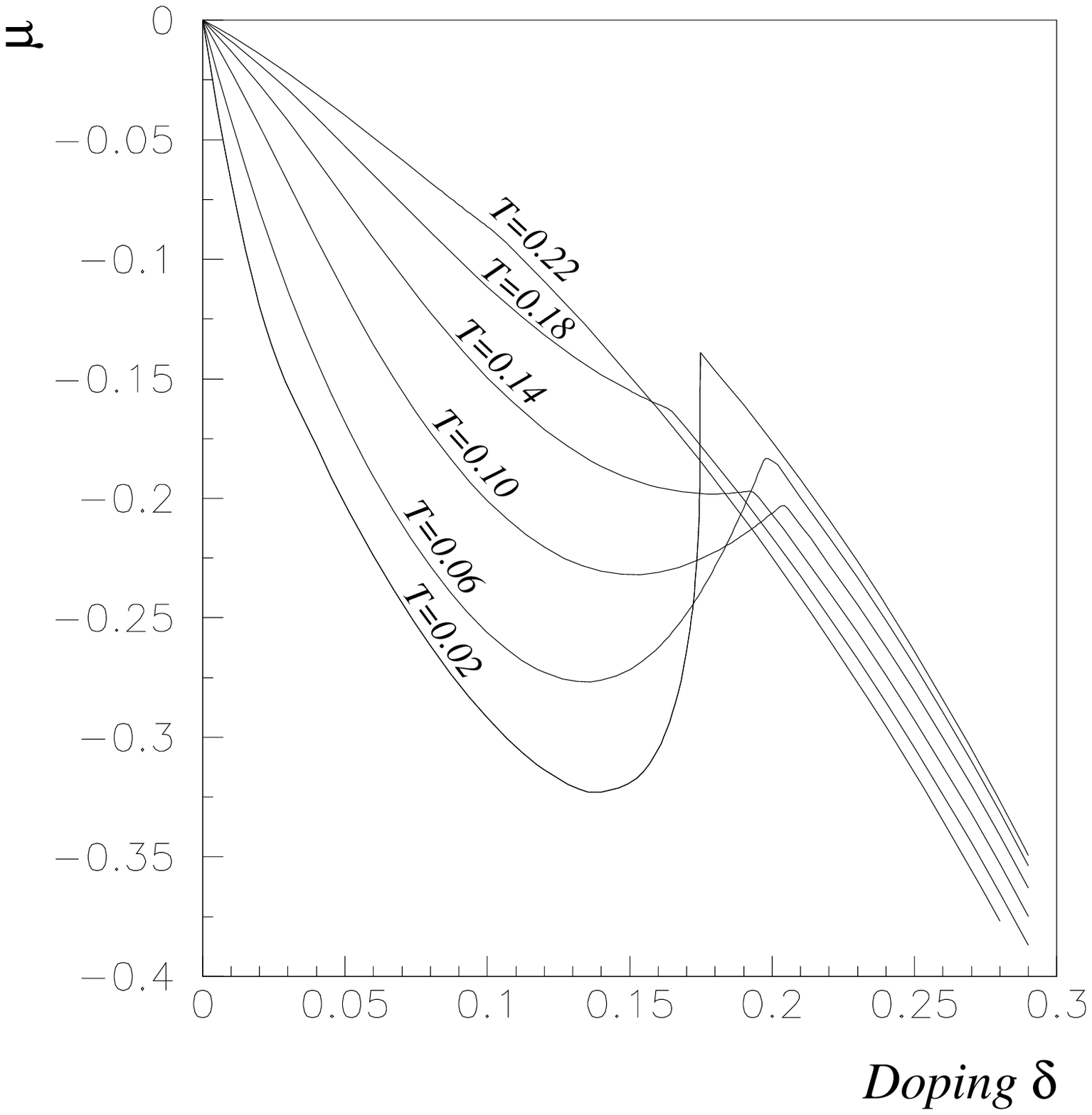}
\caption{\label{fi:chempot}}
\centerline{{\footnotesize The chemical potential $\m$ as a function
of the doping $\d$ for the flux and the uniform phases.}}
\end{center}
\end{figure}
\vskip1cm 

The function $\m(\d)$ for the flux and uniform phases
is represented in figure 5: for $\d$ below the critical
$\d_c(T)$ of the transition between the flux and the uniform, we have
plotted the chemical potential  of the flux phase, while for
$\d>\d_c(T)$ we have plotted the chemical  potential of the uniform
phase (this explains the cusp at $\d_c(T)$).  While
for $T>T_c\sim 0.145 J$ the chemical potential is always decreasing with
$\d$, when $T<T_c$ we have $\paf{\mu}{\d}>0$ in a region of doping
which extends from a certain $\d_{c0}(T)$ up to $\d_c(T)$. 
In this region the flux phase is, therefore, unstable and the system
separates in a hole-rich part, which is in the uniform phase, and in a
hole-poor part  which is in the flux phase.

The thermodynamical identities would require the free energy to be
concave only in the region of the $\d-T$ diagram with $\d_{c0}(T)<\d<
\d_c(T)$ and $T < T_c$. But we have found that, for 
temperature above $T_0 \sim 0.12 J$, the free energy of the flux
phase is already concave at $\d=0$ and remains concave up to the 
critical $\d_c(T)$. We believe that such thermodynamical inconsistency
is entirely due to some shortcoming of the mean field
approximation. In fact, let us note that the inclusion of 
thermal and quantum fluctuations might improve the consistency
between the convexity of $f(\d)$ and the monotonicity of $\mu(\d)$,
since it could substantially modify the behaviour of the free energy
while keeping the chemical potential unchanged, the latter being
uniquely determined by the saddle point equations. In the case we are
dealing with, such inclusion is, however, nontrivial and somewhat
problematic. Indeed, at $\d = \d_{c0}(T)$, one of the eigenvalues of
the fluctuation matrix becomes zero, so that the integration of
gaussian fluctuations would give actually a divergent  result. In a
similar way, special care has to be taken in the region of small $\d$:
at $\d=0$ the fluctuation matrix has, for the flux-phase, three zero
modes corresponding to the $U(1)$ gauge invariance discussed above. At
exact half filling, one can get rid of these modes  \cite{DEMTV} by
the  standard Faddev-Popov procedure. Below half filling this gauge
invariance is broken, but  for small $\d$ the eigenvalues are almost
zero and a simple gaussian integration would overestimate the
fluctuations. It is clear that next to quadratic corrections would be
required to obtain a meaningful result.

We have not found any mention to inconsistencies between the behaviour
of the free energy and that of the chemical potential in literature. In
all the previous articles discussing phase separation in the t-J
model, people have considered only one of the two approaches,  but we
believe that, at least for the works dealing with the mean field 
approximation, problems like the ones considered above would have
to be found.  
We have been able to explicitly check this in one example, 
by comparing the Maxwell contruction on the free energy which is given in 
\cite{MPP} with the behaviour of the chemical potential, 
whose analytic expression is also reported. 

Let us remark also that, at temperature $T>T_0$, we find the free 
energy to be {\it strictly} concave in the range $0\leq\d < \d_c(T)$,
which does include the zero. It is easy to see that this
problem cannot be overcome by going to negative values of the doping
(i.e. above half filling), since two phases with $\d$ of opposite sign  
cannot have the same chemical potential. Thus, in some
papers \cite{MPP,AL}, the ``bitangent construction" on the
free energy is performed by drawing the tangent to the curve $f(\d)$
from the point $(0,f(\d=0))$, forgetting the fact that this
line is not really tangent at $\d=0$. But it is not difficult to check
that this leads to considering the coexistence of two phases having
different chemical potentials, which cannot therefore be in equilibrium.

\begin{figure}[htb]
\begin{center}
\leavevmode
   \epsfxsize=0.50\textwidth\epsfbox{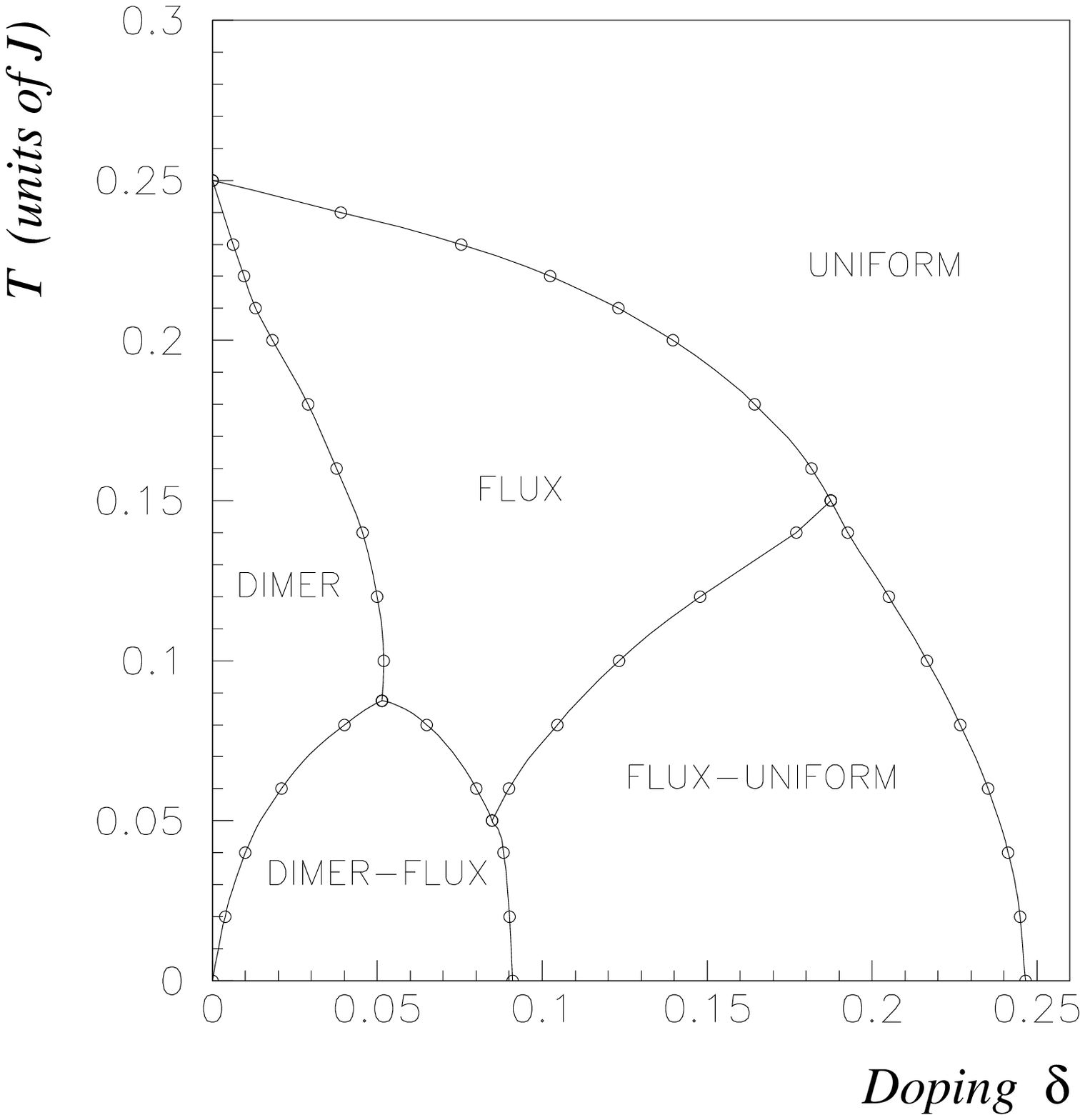}
\caption{\label{fi:phase}}
\centerline{{\footnotesize The phase diagram for the $t-J$ model,
after taking into account phase separation.}} 
\end{center}
\end{figure}
\vskip1cm 

For all the reasons explained above, we have decided to base
our analysis of phase separation on the behaviour of the  chemical
potential only, which, besides being stable against the inclusion of
fluctuations, is also in agreement with the  data on the positivity of
the eigenvalues of the fluctuation matrix. As we have already said, we
find two regions of instability, corresponding to the possibility of
coexistence between a) the flux and the uniform phases, b) the
dimer and the flux phases. The Maxwell construction on the chemical
potential and the evaluation of the free energy for the mixed phases
lead to a new phase diagram, which is substantially different from the
one of figure 4.  At high temperatures, the phase
separation instabilities disappear and hence figure 4 does not change. 
On the contrary, at low temperatures, the regions of
dimer-flux and flux-uniform phase coexistence are quite broad, and for
$T < 0.05 J$ the pure flux phase is never a minimum of the free  energy,
for any value of the doping. In addition, the kite phase is no more
present,  since it has a higher energy than the coexistence between the
flux and the uniform phase. The final phase diagram, taking into
account phase coexistence, is shown in figure 6.
Notice that we do not find receding phase boundaries anymore.

\vskip 0.3cm
{\bf Acknowledgments}

We are very grateful to G. Morandi for the many suggestions and the constant 
encouragment during the preparation of this paper. 
We also thank P. Fazekas, E. Galleani D'Agliano, F. Napoli and A. Tagliacozzo 
for helpful discussions. 

\end{document}